\documentclass[review]{elsarticle}

\usepackage{natbib}
\usepackage{lineno,hyperref}
\usepackage{amsmath}
\usepackage{bigints}

\biboptions{authoryear}

\begin{document}

\begin{frontmatter}

\title{Pathways, Scaling Laws and Analytical Solutions for Crease Formations in a Gel Layer}

\author{Xiaoyi Chen}
\author{Hui-Hui Dai\corref{mycorrespondingauthor}}
\cortext[mycorrespondingauthor]{Corresponding author}
\ead{mahhdai@cityu.edu.hk}
\address{Department of Mathematics, City University of Hong Kong,
83 Tat Chee Avenue, Hong Kong, PR China}

\begin{abstract}
An analytical study on crease formations in a swelling gel layer is conducted. By exploring the smallness of the layer thickness and using a method of coupled series-asymptotic expansions, the original nonlinear eigenvalue problem of partial differential equations is reduced to one of ordinary differential equations. The latter problem is then solved analytically to obtain closed-form solutions for all the post-bifurcation branches. With the available analytical results, a number of deep insights on crease formations are provided, including the unveiling of three pathways to crease (depending on the layer thickness), determination of the bifurcation type, establishment of a lower bound for mode numbers and two scaling laws. Also, a number of experimental results are captured, which are then nicely  interpreted based on the analytical solutions. In particular, it is shown that some critical physical quantities are invariant with respect to the thickness at the moment of crease formation.  It appears that the present work offers a comprehensive understanding on crease formation, a widely-spread phenomenon.
\end{abstract}

\begin{keyword}
\texttt{Gel}\sep Crease\sep Post-bifurcation analysis \sep Analytical method
\end{keyword}

\end{frontmatter}

\section{Introduction}
Polymeric gels are sensitive to various stimuli, such as light, temperature, force, chemical solvent and electric field\citep{Otake1990,Doi2009,TTanaka1980,TShiga1997,Zhao2008,Hong2010}. Responding to the external stimulation, gels can undergo large volume alterations. For gels subject to constraints, when the swelling or de-swelling reaches some critical condition, instabilities can occur, leading to different morphologies in the post-bifurcation regimes, e.g., buckling, wrinkling, folding and crease patterns. In particular, wrinkle and crease in gel are often observed. Wrinkles are smooth, having wavy-like shape, while creases are singular, having sharp tips at certain points. Actually, wrinkle and crease formations are widely spread phenomena, and besides gels, one can also observe wrinkling or cease patterns in our brain and skin, leaves, blossoms and rubber tyres, etc.

These patterns have attracted great research interests, and many studies have been conducted to investigate different aspects of them, e.g., critical strain for the onset of instabilities and wavelength and amplitude properties. In particular, various approaches have been used to study wrinkle formations, \citep{ZHuang2005,Huang2005,Huang2006,Im2008,Audoly2008,Cao2012,Diab2013}, including analytical treatments (linear and weakly nonlinear analysis), numerical simulations and experimental measurements. Probably, our understanding on wrinkle formations has reached a comfortable level, and people have used the knowledge in a number of applications, e.g. in flexible electric devices, tunable wetting, responsive actuators and sensors \citep{Chan2008,Yang2010,Chen2012}. On the other hand, creases were studied mainly experimentally and numerically, and here we give a brief review on some selected works.

\citet{TTanaka1987} observed experimentally the appearance of a crease pattern during the phase transition of a gel layer, which subsequently led to many further studies, see \citep{HTanaka1992,HTanaka1994,Gent1999,Ghatak2007,Trujillo2008,Hong2009,Liu2009,Hong2010,Yoon2010,Guvendiren2010,Kang2010,Jin2011,Hohlfeld2011,Cai2012,Cao2012,Yoon2012,Diab2013,Zalachas2013,Weiss2013,Chen2014,Jin2014}. \cite{Gent1999} conducted an experiment on the bending of a rubber block and pointed out that the critical strain for the onset of instability to crease was much smaller than the predicted value from Biot's linear theory \citep{Biot1963}. This indicated that crease might be a total different instability from wrinkle. \cite{Hohlfeld2011} performed a numerical simulation on the bending of a rubber block and revealed that the bifurcation to crease was nonlinear and subcritical, which obviously separated crease from wrinkle. \cite{Trujillo2008} found experimentally the effective critical strain for creasing was 0.33 for a surface-attached hydrogel, which was independent of the initial thickness of the gel layer. Later, \cite{Hong2009} found a critical strain for crease formation was 0.35 in an incompressible elastomer and deduced that the equivalent critical swelling ratio was 2.4 for a swollen gel. \cite{Cai2012} studied the formation of crease as well as its characteristic shapes both experimentally and numerically and good agreements between two approaches were found.

The transformation between wrinkle and crease is also of great interest. \cite{Guvendiren2010} showed the transition from wrinkle to crease in thin film gels could take place when the ratio of water and ethanol was changed. In studying the compression of a neo-Hookean half-space, \cite{Cao2012} showed that a wrinkling pattern with two coalescence modes  was imperfection sensitive and might be one pathway leading to crease. Recently, \cite{Diab2013} established a ruga phase diagram for different ruga  instabilities of a compressed neo-Hookean half-space with varying material property along the depth, showing that there were two pathways leading to crease, one was from wrinkle to crease (setback crease) and the other was from a homogeneous state to crease (instantaneous crease).

There are also some analytical treatments. \cite{Hwa1988} linearized the governed coupled partial differential equations (PDEs) for displacements and then analyzed the evolution of surface patterns of a swelling gel and illustrated the origin of crease using proper physical quantities. \cite{Onuki1989} studied the deformation outside the localized range of crease in a linear scheme while inside the region of crease, a simple deformation was prescribed with the region width being unknown. As nonlinearity caused by finite deformations plays an essential role in crease formations, a linearization approach may be doubtful. Notably, two
works mentioned above \citep{Cao2012,Diab2013} did consider nonlinearity in their weakly nonlinear analysis for wrinkles.
However, those results on wrinkles may not be directly used to describe creases (FEM analysis was further
used in those works). Also, \cite{Silling1991} studied the crease singularity by assuming a proper solution to the obtained PEDs for the principle stretches, but it is difficult to substantiate such an assumption.

 One drawback in numerical simulations for crease formations is that very often small localized defects at selected sites in geometry are artificially introduced to force the evolution to creases, which may limit the prediction power of this kind of approach. Also, it is difficult for a numerical scheme to capture all the post-bifurcation branches (most numerical studies only provide one branch). More desirably, some analytical descriptions are needed in order to have a more complete picture. Mathematically the instability leading to crease formation is governed by a nonlinear eigenvalue problem of complicated nonlinear PDEs. And, one needs to capture all the nonlinear eigenvalues and solutions to gain a more comprehensive understanding. Unfortunately, due to the complexity of those PDEs, so far no analytical descriptions for creases are available. Here, an attempt is made to study the instability leading to crease in a swelling polymer gel layer by an analytical method. As will be seen, the analytical results are indeed powerful, providing deep insights on crease formation as well as capturing/explaining some interesting experimental observations.
 
 One key of our success is to explore the smallness of the layer thickness to reduce the nonlinear eigenvalue problem of PDEs to that of ordinary differential equations (ODEs), which is achieved by the method of coupled series-asymptotic expansions developed by one of the authors \citep{Dai2004, Dai2006}. Luckily, we manage to construct closed-form solutions for all possible post-bifurcation branches from the reduced ODEs problem, together with a set of algebraic equations for the determination of the amplitude and induced resultant. A simple relation for the bifurcation points is also derived, leading to the finding of a lower bound for the mode number, which explains why small mode-number patterns were not observed in some experiments. With the available analytical solutions, we plot the bifurcation diagrams for all branches for three selected layer thicknesses, which show that each branch is a subcritical pitchfork bifurcation followed by a saddle-node bifurcation. The energy difference between those energy values of a post-bifurcation branch and the homogeneous state is used to judge the optimal branch (the branch with the smallest energy value) and the instability points are selected according to this branch. For a crease profile, the slope goes to infinity at the valley, which is then used to judge a wrinkle or crease pattern. It is found that, depending on the thickness, there are three pathways to creases, including one from wrinkle to crease through mode jumps, which was observed experimentally and captured here for the first time theoretically. Also, based on the deduced two scaling laws, we are able to explain several invariant quantities found in experiments.

The  sections of this paper are arranged as follows. In section 2, we give a formulation of the problem and field equations, which is followed by in section 3 a derivation of the reduced system based on the method of coupled series-asymptotic expansions. In section 4, we present the closed-form solutions to the reduced system. The bifurcation condition and  post bifurcation analysis are given in section 5, together with discussions on the results in comparisons with a number of experimental results. In section 6, we give some concluding remarks.

\section{Problem Formulation and Field Equations}

\subsection{Problem Formulation}
\begin{figure}
\begin{center}
\includegraphics[scale=0.7]{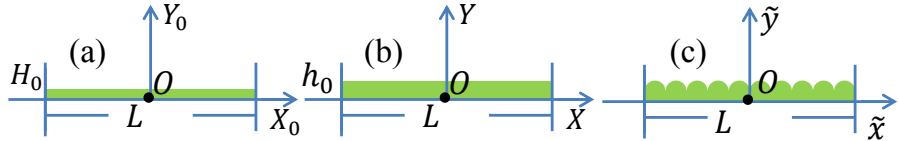}
\caption{\label{Fig_1} Sketch of the swelling states (a)$B_0$, (b) $B^*$ and (c) $B_t$.}
\end{center}
\end{figure}

We consider a gel layer immersed in a solvent, which is constrained horizontally by two lubricated rigid walls and vertically by a lubricated rigid substrate (or a very stiff Winkler foundation such that there are no vertical displacement and shear stress) and with no constraint in the third direction. Fig. \ref{Fig_1} is a schematic description of the problem. Initially, the gel has a thickness $H_0$ and a length $L$ and the configuration is denoted by $B_0$. Then, mixing with solvent molecules, the gel can swell vertically to a homogeneous state $B^*$ with height $h_0=\lambda_2 H_0 $ where $\lambda_2$ is the vertical stretch, which is regarded as the loading parameter. Due to instability, an inhomogeneous state can also develop, which is denoted by $B_t$. We use $(X^0, Y^0)$, $(X, Y)$ and $(\tilde{x}, \tilde{y})$ to denote a material point in states $B_0$, $B^*$ and $B_t$ respectively. It can be seen that $X=X^0, Y=\lambda_2 Y^0$, and then the two incremental displacements superimposed on $B^*$ are given by $U(X, Y)=\tilde{x}(X, Y)-X, V(X, Y)=\tilde{y}(X, Y)-Y$.
We adopt the free energy density function introduced by \citet{Flory1943}, which takes the following form upon using the molecular incompressibility \citep{Treolar1975,Pence2005} to replace the solvent concentration (see \cite{Liu2009}):
\begin{equation}
\label{Eq1}
\begin{array}{l}
\phi(\textbf{F},\mu)=\frac{1}{2}N \kappa T(I-3-2\ln{J})-\frac{\kappa T}{v}(J-1)\ln(\frac{J}{J-1})-\frac{\kappa T}{v}\frac{\chi}{J}-\frac{\mu}{\nu}(J-1),
\end{array}
\end{equation}
where $\textbf{F}$ is the deformation gradient, $I=F_{ik}F_{ik}$
and $J^2=(\textrm{det}~\textbf{F})^2$.
And, $T$ and $\mu$ are respectively the temperature and chemical
potential of the solvent, $N$ is the
number of polymer chains per reference volume of dry network,
$\nu$ is the volume per solvent molecule, $\kappa$ is the Boltzmann's constant and $\chi$ is the dimensionless measure of the enthalpy of mixing. In the sequel, $\kappa T=4\times 10^{-21} \mathrm{J}$, $\nu=10^{-28} \mathrm{m}^3$, $N\nu=10^{-3}$ and $\chi=0.4$ will be used. We note that $\mu$ can be calculated for a prescribed $\lambda_2$, and vice versa.

\subsection{Field Equations}
Taking $B^*$ as the reference state, the field equations and boundary conditions are (see \citet{Fu1999})
\begin{eqnarray}
\label{Eq2}
\begin{array}{l}
\frac{\partial{\chi_{11}}}{\partial{X}}+\frac{\partial{\chi_{12}}}{\partial{Y}}=0,~~~~~~~~~~~\chi_{12}=0,~~V(X, Y)=0,~~~\textrm{at}~~ Y=0,

\\
~~~~~~~~~~~~~~~~~~~~~~~~~~\textrm{and}~~~~\chi_{12}=0,~~\chi_{22}=0, ~~~~~~~~\textrm{at}~~Y=h_0,
\\
\frac{\partial{\chi_{21}}}{\partial{X}}+\frac{\partial{\chi_{22}}}{\partial{Y}}=0,~~~~~~~~~~~\chi_{21}=0,~~U(X, Y)=0,~~~\textrm{at}~~ X=\pm \frac{L}{2},
\end{array}
\end{eqnarray}
where $\chi_{ij}=(S_{iA}-\bar{S}_{iA})\bar{F}_{jA}$ ($i,j, A=1,2$), $\bar{F}_{jA}$ are the components of the deformation gradient at state $B^*$ and $S_{iA}$ and $\bar{S}_{iA}$ the components of the first Piola-Kirchhoff stress at states $B^*$ and $B_t$ respectively. This system represents a nonlinear eigenvalue problem for finding nontrivial solutions for two unknowns $U(X, Y)$ and $V(X, Y)$ with eigenvalue $\lambda_2$. It appears that there lack analytical tools to analyze this complex system of PDEs directly. Here, we use the method of coupled series-asymptotic expansions (see \citet{Dai2004,Dai2006}) to do a dimension reduction to eliminate the variable $Y$, taking advantage of the smallness of $h_0$ ($\frac{h_0}{L}$ is assumed to be very small).

For a weakly nonlinear analysis, we expand $\chi_{ij}$ for small strains up to the third order,
\begin{equation}
\label{Eq3}
\chi_{ij}=a_{jilk}^1d_{kl}+\frac{1}{2}a_{jilknm}^2d_{kl}d_{mn}+\frac{1}{6}a_{jilknmqp}^3d_{kl}d_{mn}d_{pq}+O(|d_{ij}|^4),
\end{equation}
where $d_{ij}=\frac{\partial{D_i}}{\partial{X^j}}$ (here for convenience we denote $D_1=U(X, Y), D_2=V(X, Y), X^1=X, X^2=Y$) are the components of the displacement gradient tensor and $a_{jilk}^1,~a_{jilknm}^2$, $a_{jilknmqp}^3$ are elastic moduli defined by
\begin{equation}
\label{Eq4}
\begin{aligned}
&a_{jilk}^1=\bar{F}_{jA}\bar{F}_{lB}\frac{\partial{^2\phi}}{\partial{F_{iA}}\partial{F_{kB}}}|_{F=\bar{F}},\\
&a_{jilknm}^2=\bar{F}_{jA}\bar{F}_{lB}\bar{F}_{nC}\frac{\partial{^3\phi}}{\partial{F_{iA}}\partial{F_{kB}}\partial{F_{mC}}}|_{F=\bar{F}},\\    &a_{jilknmqp}^3=\bar{F}_{jA}\bar{F}_{lB}\bar{F}_{nC}\bar{F}_{tG}\frac{\partial{^4\phi}}{\partial{F_{iA}}\partial{F_{kB}}\partial{F_{mC}}\partial{F_{sG}}}|_{F=\bar{F}},
\end{aligned}
\end{equation}
where $\bar{F}_{ij}$ are the components of $\bar{\mathbf{F}}$, which is the deformation gradient from $B_0$ to $B^*$ where the height of the gel layer reaches $h_0$. From the relationship of $(X^0, Y^0)$ and $(X, Y)$ presented in the previous subsection, we have
\begin{equation*}
\mathbf{\bar{F}}= \left( \begin{array}{ccc} 1 &  & \\ & \lambda_2 &  \\
 &  & \lambda_2
\end{array}
\right),
\end{equation*}
where $\lambda_2$ can be obtained by setting the first Piola-Kirchhoff stress component $\bar{S}_{22}=0$ (which gives a relation between $\lambda_2$ and $\mu$). Note that  $a_{jilk}^1$,~$a_{jilknm}^2$, $a_{jilknmqp}^3$ can be calculated by using the formulas given in \cite{Fu1999} or directly by MATHEMATICA for the previously given $\phi$. After substituting Eq. $(\ref{Eq3})$ into Eq. $(\ref{Eq2})$, one has a coupled PDEs with cubic nonlinearity, which are still very complicated.

\section{Reduced System Based on Coupled Series-asymptotic Expansions}

\subsection{Nondimensionalization}
To do the reduction to eliminate the $Y$ variable, we introduce the following scalings to non-dimensionalize Eq. $(\ref{Eq2})$:
\begin{equation}
\label{Eq5}
\begin{aligned}
\small
&x=\frac{X}{L},~~~~y=\frac{Y}{L},~~~~\epsilon=\frac{h}{L},~~~~\eta=\frac{h_0}{L},\\
&u(x, y)=\frac{U(X, Y)}{h},~~~~v(x, y)=\frac{V(X, Y)}{h},
\end{aligned}
\end{equation}
where $x$ and $y$ are the dimensionless coordinates with $-\frac{1}{2} \leq x \leq \frac{1}{2}$ and $0 \leq y \leq \frac{h_0}{L}=\eta$, and $h$ is the characteristic magnitude of the vertical displacement and $\epsilon$ is regarded to be a small parameter. The layer is thin and thus $\eta$ is small, which implies $y$ varying in a small interval.
The dimensionless displacements $u(x, y)$ and $v(x, y)$  are assumed to be smooth enough in $y$ and thus can be expanded about $y=\eta$ as series
\begin{equation}
\label{Eq6}
\begin{aligned}
&u(x, y)= \delta (U_0(x)+(\eta-y)^2 U_2(x)+\cdots)+((\eta-y) U_1(x)+(\eta-y)^3 U_3(x)+\cdots),\\
&v(x, y)= (V_0(x)+(\eta-y)^2 V_2(x)+\cdots)+\delta ((\eta-y) V_1(x)+(\eta-y)^3 V_3(x)+\cdots),
\end{aligned}
\end{equation}
where $\delta$ is a parameter and the introduction of $\delta$ allows the horizontal displacement $U(X, Y)$ to have a different magnitude ($\delta h$) from that ($h$) of the vertical displacement $V(X, Y)$. Noteworthily,  we do not assume the smallness of $\delta$, instead $O(\epsilon \eta^3\delta, \epsilon^2 \eta^3\delta^2, \epsilon^3\eta^2\delta^2) $ terms are assumed to be small (cf. Eq. $(\ref{Eq7})$).

\subsection{Reduced Equations}
Substituting Eqs. $(\ref{Eq3})$-$(\ref{Eq6})$  into the boundary conditions at the top and bottom surfaces in Eq. $(\ref{Eq2})$, we obtain
\begin{equation}
\label{Eq7}
\begin{aligned}
~~~~~~~~~~~~~0&=\chi_{12}|_{Y=0}\\
&=\epsilon (c_1 U_1+c_2 V_0^{\prime})+\epsilon \eta \delta(c_3 U_2+c_4 V_1^{\prime})+\epsilon \eta^2(c_5 U_3+c_6 V_2^{\prime})\\
&~~~~+\epsilon^2\delta(c_7 V_1 V_0^{\prime}
+c_8 U_0^{\prime} V_0^{\prime})+\epsilon^2\eta(c_9 V_2 V_0^{\prime}+c_{10} U_1^{\prime} V_0^{\prime})\\
&~~~~+\epsilon^2\eta \delta^2(c_{11} V_1 V_1^{\prime}+C_{12} U_0^{\prime} V_1^{\prime})
+\epsilon^2\eta^2\delta(c_{13} V_3 V_0^{\prime}+c_{14} U_2^{\prime} V_0^{\prime}\\
&~~~~+c_{15} V_2 V_1^{\prime}+c_{16} U_1^{\prime} V_1^{\prime}+c_{17} V_1 V_2^{\prime}
+c_{18} U_0^{\prime} V_2^{\prime})+\epsilon^3\delta^2(c_{19} V_1^2 V_0^{\prime}\\
&~~~~+c_{20} V_1 U_0^{\prime} V_0^{\prime}+c_{21} {U_0^{\prime}}^2 V_0^{\prime})+\epsilon^3 c_{22} U_1 {V_0^{\prime}}^2+\epsilon^3\eta \delta(c_{23} V_1 V_2 V_0^{\prime}\\
&~~~~+c_{24} V_2 U_0^{\prime} V_0^{\prime}+c_{25} V_1 U_1^{\prime} V_0^{\prime}+
+c_{26} U_0^{\prime} U_1^{\prime} V_0^{\prime}+c_{27} U_2 {V_0^{\prime}}^2+c_{28} U_1 V_0^{\prime} V_1^{\prime})\\
&~~~~+\epsilon^3\eta \delta^3(c_{29} V_1^2 V_1^{\prime}+c_{30} V_1 V_1^{\prime} U_0^{\prime}+c_{31} V_1^{\prime} {U_0^{\prime}}^2)
+O(\epsilon \eta^3\delta, \epsilon^2 \eta^3\delta^2, \epsilon^3\eta^2\delta^2),\\
0&=\frac{V(X, Y)|_{Y=0}}{L}=\epsilon V_0+\epsilon \eta  \delta V_1+\epsilon \eta^2 V_2+O(\epsilon \eta^3\delta, \epsilon^2 \eta^3\delta^2, \epsilon^3\eta^2\delta^2),\\
0&=\chi_{12}|_{Y=h_0}\\
&=\epsilon(c_{32} U_1+c_{33} V_0^{\prime})+\epsilon^2 \delta(c_{34} V_1 V_0^{\prime}+c_{35} U_0^{\prime} V_0^{\prime})+\epsilon^3 C_{36} U_1 {V_0^{\prime}}^2\\
&~~~~+\epsilon^3\delta^2(c_{37} V_1^2V_0^{\prime}
+c_{38} V_1 U_0^{\prime} V_0^{\prime}+c_{39} {U_0^{\prime}}^2 V_0^{\prime}),\\
0&=\chi_{22}|_{Y=h_0}\\
&=\epsilon \delta(c_{40} V_1+c_{41} U_0^{\prime})+\epsilon^2 c_{42} U_1 V_0^{\prime}+\epsilon^2\delta^2(c_{43} V_1^2+c_{44} V_1 U_0^{\prime} +c_{45} {U_0^{\prime}}^2)\\
&~~~~+\epsilon^3 \delta(c_{46} U_1 V_1 V_0^{\prime}+c_{47} U_1 U_0^{\prime} V_0^{\prime})+\epsilon^3\delta(c_{48} V_1^3+c_{49} V_1^2 U_0^{\prime}+c_{50} V_1 {U_0^{\prime}}^2\\
&~~~~+c_{51} {U_0^{\prime}}^3), 
\end{aligned}
\end{equation}
where a {\it prime} denote the derivative with respect to $x$. Here and in the sequel $c_i$ are coefficients depending on $\lambda_2$, whose lengthy expressions are omitted. Similar to the treatment in \cite{Dai2013}, in order to obtain a closed system for a finite number of unknowns in an asymptotically consistent manner, we neglect $O(\epsilon \eta^3 \delta, \epsilon^2 \eta^3 \delta^2, \epsilon^3 \eta^2 \delta^2)$ terms in Eqs. $(\ref{Eq7})_1$ and  $(\ref{Eq7})_2$. Then the above system of four equations contains eight unknowns: $U_i(x)$ and $V_i(x)$ ($i=0, 1, 2, 3$). Another four equations are needed, which can be obtained from the field equations.

Substituting Eqs. $(\ref{Eq3})$-$(\ref{Eq6})$ into the field equations in Eq. $(\ref{Eq2})$, the left-hand sides become two infinite series of $(\eta-y)$, and all the coefficients of $(\eta-y)^n~(n=0,1, \cdots)$ should be zero. It turns out that the coefficients of $(\eta-y)^0$ and $(\eta-y)^1$ contain and only contain the above-mentioned eight unknowns, and as a results we have the following four equations:
\begin{equation}
\label{Eq8}
\begin{aligned}
&t_0(\mathbf{w_1})+t_1(\mathbf{w_1})\cdot U_2+t_2(\mathbf{w_1})\cdot V_2=0,\\
&g_0(\mathbf{w_1})+g_1(\mathbf{w_1})\cdot U_2+g_2(\mathbf{w_1})\cdot V_2=0,  \\
&t_3(\mathbf{w_2})+t_4(\mathbf{w_2})\cdot U_3+t_5(\mathbf{w_2})\cdot V_3=0,  \\
&g_3(\mathbf{w_2})+g_4(\mathbf{w_2})\cdot U_3+g_5(\mathbf{w_2})\cdot V_3=0,
\end{aligned}
\end{equation}
where $\mathbf{w_1}=(U_0, V_0, U_1, V_1),~\mathbf{w_2}=(U_0, V_0, U_1, V_1, U_2, V_2)$, and $t_i, g_j (i,j=0, \cdots, 5)$ are polynomial functions (also containing parameters $\epsilon$ and $\delta$), whose expressions are omitted. It can be observed that Eqs. $(\ref{Eq8})_1$ and $(\ref{Eq8})_2$ are two linear algebraic equations for $U_2$ and $V_2$ while Eqs. $(\ref{Eq8})_3$ and $(\ref{Eq8})_4$ are linear algebraic equations for $U_3$ and $V_3$. Solving them and further using asymptotic expansions in the small parameter $\epsilon$, we obtain
\begin{equation}
\label{Eq9}
\begin{aligned}
U_2=&c_{64} V_1^{\prime}+c_{65}U_0^{\prime\prime}+\frac{\epsilon}{\delta}(c_{66} U_1^{} V_0^{}+c_{67} U_1 V_0^{\prime\prime}+c_{68} V_0^{\prime} V_0^{\prime\prime})+\epsilon \delta(c_{69} V_1 V_1^{\prime}+c_{70} U_1 V_0^{\prime\prime}\\
&+c_{71} U_0^{\prime} V_1^{\prime}+c_{72} U_0^{\prime} U_0^{\prime\prime})
+\epsilon^2(c_{73} V_0^{\prime} V_1^{\prime} U_1+c_{74} V_0^{\prime}U_0^{\prime\prime} U_1+c_{75} {V_0^{\prime}}^2 V_1^{\prime}+c_{76} {V_0^{\prime}}^2 U_0^{\prime\prime}\\
&+c_{77} U_0^{\prime} U_1^{\prime} V_0^{\prime}+c_{78} U_0^{\prime} U_1 V_0^{\prime\prime}+c_{79} U_0^{\prime} V_0^{\prime} V_0^{\prime\prime}+c_{48}U_1^{\prime} V_0^{\prime} V_1+c_{80} U_1 V_0^{\prime\prime} V_1\\
&+c_{81}V_0^{\prime} V_0^{\prime\prime} V_1)+\epsilon^2\delta^2(c_{83} V_1 V_1^{\prime} U_0^{\prime}+c_{84} U_0^{\prime} U_0^{\prime\prime} V_1+c_{85} V_1^2 V_1^{\prime}+c_{86} V_1^2 U_0^{\prime\prime}\\
&+c_{87} {U_0^{\prime}}^2 V_1^{\prime}+c_{88} {U_0^{\prime}}^2 U_0^{\prime\prime})+O(\epsilon^3\delta^3),\\
V_2=&c_{89} U_1^{\prime}+c_{90} V_0^{\prime\prime}+\epsilon \delta(c_{91} U_1 V_1^{\prime}+c_{92} V_0^{\prime} V_1^{\prime}+c_{93} U_1 U_0^{\prime\prime}+c_{94} V_0^{\prime} U_0^{\prime\prime}+c_{95} V_1 U_1^{\prime}\\
&+c_{96} V_1 V_0^{\prime\prime}+c_{97} U_0^{\prime} U_1^{\prime}
+c_{98} U_0^{\prime} V_0^{\prime\prime})+O(\epsilon^2\delta^2),\\
U_3=&c_{99} V_2^{\prime}+c_{100} U_1^{\prime\prime}+\epsilon \delta(c_{101} U_2^{\prime} V_0^{\prime}+c_{102} V_2 V_1^{\prime}+c_{103} U_1^{\prime} V_1^{\prime}+c_{104} V_1 V_2^{\prime}+c_{105} U_0^{\prime} V_2^{\prime}\\
&+c_{106} V_2 U_0^{\prime\prime}+c_{107} U_1^{\prime} U_0^{\prime\prime}
+c_{108} V_1 U_1^{\prime\prime}+c_{109} U_0^{\prime} U_1^{\prime\prime}+c_{110} U_2 V_0^{\prime\prime}+c_{111} U_1 V_1^{\prime\prime}\\
&+c_{112} V_0^{\prime} V_1^{\prime\prime})+O(\epsilon^2\delta^2),\\
V_3=&c_{113} U_2^{\prime}+c_{114} V_1^{\prime\prime}+O(\epsilon \delta).
\end{aligned}
\end{equation}
After substituting Eq. (\ref{Eq9}) into Eq. (\ref{Eq7}), we have only four equations for four unknowns $(U_0, U_1, V_0, V_1)$. It turns out that further simplifications are possible.

We notice that Eqs. $(\ref{Eq7})_3$ and $(\ref{Eq7})_4$ are exact equations, and particularly Eq. $(\ref{Eq7})_3$ is a linear algebraic equation for $U_1$. As a result, upon further using the smallness of $\epsilon$, we obtain
\begin{equation}
\label{Eq10}
\begin{aligned}
U_1&=c_{52} V_0^{\prime}+\epsilon \delta(c_{53} V_1 V_0^{\prime} +c_{54} U_0^{\prime} V_0^{\prime})+\epsilon^2\delta^2(c_{55} V_1^2 V_0^{\prime}+c_{56} V_1 U_0^{\prime} V_0^{\prime}+c_{57} {U_0^{\prime}}^2 V_0^{\prime}) \\
&~~~~+\epsilon^2 c_{58} {V_0^{\prime}}^3+O(\epsilon^3\delta).
\end{aligned}
\end{equation}
The substitution of Eq. (\ref{Eq10}) into Eq. $(\ref{Eq7})_4$ yields an equation containing $(U_0, V_0, V_1)$, which turns out to be a cubic equation for $V_1$. By using the smallness of $\epsilon$ and a regular perturbation method \citep{Bush1992,Holmes2013}, this equation is solved to give
\begin{equation}
\label{Eq11}
V_1=c_{59} U_0^{\prime}+\epsilon \delta c_{60} {U_0^{\prime}}^2+\frac{\epsilon}{\delta}c_{61} {V_0^{\prime}}^2+\epsilon^2\delta^2c_{62} {U_0^{\prime}}^3+\epsilon^2 c_{63} U_0^{\prime} {V_0^{\prime}}^2+O(\epsilon^3\delta^3).
\end{equation}

Eliminating $U_1$ and $V_1$ in Eqs. $(\ref{Eq7})_1$ and $(\ref{Eq7})_2$ by using the above two equations, we obtain the following two nonlinear coupled ODEs for $U_0$ and $ V_0$:
\begin{equation}
\label{Eq12}
\begin{aligned}
C&=\epsilon \eta \delta  c_{115} U_0^{\prime}+\epsilon \eta^2 c_{116} V_0^{\prime\prime}+\epsilon^2\eta c_{117} {V_0^{\prime}}^2+\epsilon^2\eta\delta^2 c_{118} {U_0^{\prime}}^2+\epsilon^2\eta^2\delta c_{119} U_0^{\prime} V_0^{\prime\prime}\\
&~~~~+\epsilon^3 \eta  \delta c_{120} U_0^{\prime} {V_0^{\prime}}^2
+\epsilon^3 \eta  \delta^3c_{121} {U_0^{\prime}}^3+O(\epsilon \delta \eta^3, \epsilon^2\delta^2\eta^3, \epsilon^3\delta^2\eta^2),\\
0&=\epsilon V_0+\epsilon \eta \delta c_{122} U_0^{\prime}+\epsilon \eta^2 c_{123} V_0^{\prime\prime}+\epsilon^2 \eta c_{124} {V_0^{\prime}}^2+\epsilon^2 \eta  \delta^2 c_{125} {U_0^{\prime}}^2\\
&~~~~+\epsilon^2 \eta^2 \delta c_{126} U^{\prime}_0 V^{\prime \prime} _0 +\epsilon^3 \eta \delta  c_{127} U_0^{\prime} {V_0^{\prime}}^2
+\epsilon^3 \eta \delta^3 c_{128} {U_0^{\prime}}^3+O(\epsilon \delta \eta^3, \epsilon^2\delta^2\eta^3, \epsilon^3\delta^2\eta^2).
\end{aligned}
\end{equation}
Note that Eq. $(\ref{Eq12})_1$ is obtained after a further integration and $C$ is the integration constant. It can be shown that $\frac{C L}{h_0}=\frac{1}{H_0}\bigintss_0^{H_0} \chi_{11} \mathrm{d} Y_0=:P$ ($P$ is the average incremental resultant in the horizontal direction). A further observation is that Eq. $(\ref{Eq12})_1$ is a cubic equation for $U_0^{\prime}$, which can be solved by using a regular perturbation method with two small parameters $\epsilon$ and $\eta$. As a result, upon retaining the original variables with $U(X,0)=W(X)$ and $V(X,0)=G(X)$, we obtain
\begin{equation}
\label{Eq13}
\begin{aligned}
0&=G(X)+(E_{1}+f_1(P)) \lambda_2 H_0 (G'(X))^2+(E_{2}+f_2(P))(\lambda_2 H_0)^2 G^{\prime \prime}(X)+\lambda_2 H_0 f_3(P),\\
&W^{\prime}(X)=f_4(P) \lambda_2 H_0 ~ G^{\prime\prime}(X)+ f_5(P) (G^{\prime}(X))^2+f_6(P),
\end{aligned}
\end{equation}
where a $prime$ denotes the derivative with respect to $X$ and the coefficients $f_i(P) (i=1,\cdot\cdot\cdot,6)$ depend on both $P$ and $\lambda_2$ (noteworthily $f_3(0)=0, f_6(0)=0$) and $E_1$ and $E_2$ depend on $\lambda_2$ only. So, the problem reduces to solving these two nonlinear ODEs.

\subsection{Reduced Boundary Conditions}
Now, we derive proper boundary conditions for the reduced system. Upon using Eqs. (\ref{Eq3})-(\ref{Eq7}), (\ref{Eq9})-(\ref{Eq11}) in boundary conditions at $X=\pm \frac{L}{2}$ in Eq. $(\ref{Eq2})$, we have
\begin{equation}
\label{Eq14}
\begin{aligned}
&0=\chi_{21}\\
&~~=G^{\prime}(X) \big(c_{148}+c_{149} W^{\prime}(X)+c_{150} (W^{\prime}(X))^2+c_{151} (G^{\prime}(X))^2 \big)+(h_0-Y)\\
&~~~~~~\Big(G^{\prime}(X) \big(c_{152} G^{\prime \prime}(X)
+c_{153}G^{\prime}(X) W^{\prime \prime}(X)+c_{154} G^{\prime \prime}(X) W^{\prime}(X)\big)+W^{\prime \prime}(X) \\
&~~~~~~\big(c_{155}+c_{156} W^{\prime}(X)+c_{157}(W^{\prime}(X))^2\big)\Big),~~~\mathrm{at}~~~X=\pm \frac{L}{2},\\
&0=U(X, Y)=W(X)+(h_0-Y) G^{\prime}(X) (c_{144}+c_{145} W^{\prime}(X)+c_{146} (W^{\prime}(X))^2\\
&~~~~~~+c_{147} (G^{\prime}(X))^2),~~~\mathrm{at}~~~X=\pm \frac{L}{2}.
\end{aligned}
\end{equation}
Further, from Eq. $(\ref{Eq13})$ one can obtain
\begin{equation}
\label{Eq15}
\begin{aligned}
W^{\prime \prime}(X)&=G^{\prime}(X) (f_4(P) (1+2 (E_1+f_1(P)) \lambda_2 H_0 G^{\prime \prime}(X))\\
&~~~~(-(E_2+f_2(P)) \lambda_2 H_0)^{-1}+2 f_5(P) G^{\prime \prime}(X) ).
\end{aligned}
\end{equation}
Substituting the above equation into Eq. $(\ref{Eq14})$, it is easy to deduce
\begin{equation}
\label{Eq16}
\begin{aligned}
&G^{\prime}(X)=0, ~~~~at~~~~X=\pm \frac{L}{2},\\
&W(X)=0,~~~~at~~~~X=\pm \frac{L}{2}.
\end{aligned}
\end{equation}
Eqs. (\ref{Eq13}) and (\ref{Eq16}) compose the reduced nonlinear eigenvalue problem of ODEs.  In the sequel,  without loss of generality, we set $L=1$.

\section{Solutions to the Reduced System}

\subsection{Closed-form Solutions}
A simple phase-plane ($G-G^{\prime}$ plane) analysis on Eq. $(\ref{Eq13})$ shows that for bounded $G$ the only possible solutions are periodic ones (for which $G'$ is also bounded). Luckily, Eq. $(\ref{Eq13})_1$ can be integrated twice. By introducing ${\tilde{G}=[G+\lambda_2 H_0 f_3(P)]{H^{-1}_0}}$,  Eq. $(\ref{Eq13})_1$ can be rewritten as 
\begin{equation}
0=(\tilde{G} H_0)+(E_1+f_1(P)) \lambda_2 H_0 (\tilde{G}^{\prime}(X) H_0)^2+\frac{1}{2} (E_2+f_2(P)) (\lambda_2 H_0)^2 \frac{\mathrm{d} (\tilde{G}^{\prime} (X) H_0)^2 }{\mathrm{d} (\tilde{G} H_0)},\nonumber
\end{equation}
which can be integrated once to give
\begin{eqnarray}
\label{Eq17}
\begin{array}{l}
{(H_0\tilde{G}^{\prime}(X))}^2= F(\tilde{G}),~~~F(\tilde{G})=\frac{-\tilde{G}}{D_1} +\frac{D_2}{2 D_1^2}+C_1\exp({\frac{-2 D_1\tilde{G}}{D_2} }),
\end{array}
\end{eqnarray}
where $C_1$ is an integration constant and the coefficients $D_1$ and $D_2$ depend on both $P$ and $\lambda_2$.  Then,  a further integration leads to the following expressions for the periodic solutions
\begin{eqnarray}
\label{Eq18}
\begin{array}{l}
X+(\frac{1}{2}-\frac{m}{2 k})=\int_{G_1}^{\tilde{G}(X)}\frac{H_0}{\sqrt{F(\tilde{G})}} \mathrm{d}\tilde{G},\\
\textrm{for}~~-\frac{1}{2}+\frac{m}{2 k}\leq X \leq -\frac{1}{2}+\frac{m+1}{2 k},~~~~m=0,2,\cdots, 2k-2,~~
\textrm{and}
\\
\\
X+(\frac{1}{2}-\frac{m}{2 k})=\int_{G_2}^{\tilde{G}(X)}\frac{H_0}{-\sqrt{F(\tilde{G})}} \mathrm{d}\tilde{G},\
\\
\textrm{for}~~-\frac{1}{2}+\frac{m}{2 k}\leq X \leq -\frac{1}{2}+\frac{m+1}{2 k},~~~~m=1,3,\cdots, 2k-1,
\end{array}
\end{eqnarray}
where $G_1= \tilde{G}(\frac{-1}{2})$, $G_2=\tilde{G} (\frac{-1}{2}+\frac{1}{2 k})$ and $k$ is the wave number. Two scaling laws can also be observed. Under the scalings of $[X+(\frac{1}{2}-\frac{m}{2 k})]H^{-1}_0 \rightarrow X_1$ or $[X+(\frac{1}{2}-\frac{m}{2 k})]H^{-1}_0\rightarrow X_2$ for an even or odd $m$, we have $\tilde{G}=\tilde{G}(X;\lambda_2, H_0)= \tilde{G}(X_1;\lambda_2)~{\rm or}~\tilde{G}(X_2;\lambda_2)$. And by setting $\tilde{W}=W H_0^{-1}$, from Eq. $(\ref{Eq13})_2$ we get
$W^{\prime}(X;\lambda_2, H_0)=\tilde{W}^{\prime}(X_1, \lambda_2)$
 or $\tilde{W}^{\prime}(X_2, \lambda_2)$. Later on, we shall see that these scaling laws have some important implications.

\subsection{Determination of the Parameters}
 The above obtained expressions for nontrivial solutions contain parameters $G_1, G_2, P$ and $C_1$, which can be determined by using end boundary conditions and periodicity.
 
 Since $\tilde{G}^{\prime}(X)|_{X=\pm \frac{1}{2}}=0$, $G_1$ is either the minimum or maximum in one period (so does $G_2$). Without loss of generality, we take $G_1$ as the minimum and then $G_2$ is the maximum. Upon using $F(G_1)=0$ and $F(G_2)=0$, $C_1$ can be represented in terms of $G_1$ or $G_2$, and further we have
\begin{eqnarray}
\label{Eq19}
\begin{array}{l}
0=-\frac{G_2}{D_1}+\frac{D_2}{2 D_1^2}+(\frac{G_1}{D_1}-\frac{D_2}{2 D_1^2})\exp{\frac{2 D_1}{D_2} (G_1-G_2)}.
\end{array}
\end{eqnarray}
Also, in a half period the interval length should be $\frac{1}{2k}$, and as a result from Eq. (\ref{Eq18}) we have
\begin{eqnarray}
\label{Eq20}
\begin{array}{l}
1/(2 \tilde{k})=\int_{G_1}^{G_2}\frac{1}{\sqrt{F(\tilde{G})}} \mathrm{d}\tilde{G},
\end{array}
\end{eqnarray}
where $\tilde{k}= k H_0$. Further, from the boundary conditions for $W$ one can deduce
\begin{eqnarray}
\label{Eq21}
\begin{array}{l}
0=\int_{G_1}^{G_2}\frac{\tilde{G}-E}{\sqrt{F(\tilde{G)}}} \mathrm{d}\tilde{G},
\end{array}
\end{eqnarray}
where $E=\frac{f_6(P) D_1}{f_5(P)}$. When the loading parameter $\lambda_2$ is prescribed, Eqs. $(\ref{Eq19})$-$(\ref{Eq21})$ can be used to find $G_1, G_2, P$ for each mode $k$ (or a value of $\tilde{k}$ for a given $H_0$). Once the periodic solution for $G$ is obtained from Eq. $(\ref{Eq18} )$, the solution for $W$ can be found from Eq. $(\ref{Eq13})_2$.

\subsection{The Total Energy}
Now, we shall give the expression for the total energy, which will be needed for identifying the optimal branch (the branch with the smallest energy value).

We point out that with those results derived above, all strain components can be represented in terms of $\tilde{G}(=[G+\lambda_2 H_0 f_3(P)]H_0^{-1})$ (according to the scaling laws given previously $\tilde{G}$ is independent of $H_0$). The free energy density function is given in Eq. (\ref{Eq1}), and we denote the dimensionless free energy (scaled by the modulus $N \kappa T$) in state $B_t$ by $\phi_t$ and its through-thickness average by $\bar{\phi}_t$ ($=\frac{1}{H_0}\bigintss_{0}^{H_0} \phi_t \mathrm{d}Y$). As $\bar{\phi}_t$ is a function of strains,  it can also be represented in terms of $\tilde{G}$, and the expression is
\begin{eqnarray}
\label{Eq22}
\begin{array}{l}
\bar{\phi}_t=h_{1}(P)+h_{2}(P) F(\tilde{G}) +h_{3}(P) F^2(\tilde{G})+h_{4}(P) F^3(\tilde{G})+h_{5}(P) F^4(\tilde{G})\\
~~~~~~+\tilde{G} (h_{6}(P)+h_{7}(P) F(\tilde{G})
+h_{8}(P) F^2(\tilde{G})+h_{9}(P) F^3(\tilde{G}))+\tilde{G}^2 (h_{10}(P)\\
~~~~~~+h_{11}(P) F(\tilde{G}) +h_{12}(P) F^2(\tilde{G}))+\tilde{G}^3 (h_{13}(P)
+h_{14}(P) F(\tilde{G}))+h_{15}(P) \tilde{G}^4,
\end{array}
\end{eqnarray}
where the coefficients $h_i(P)$ depend on both $P$ and $\lambda_2$. Then, the total energy scaled by $H_0$ is
\begin{eqnarray}
\label{Eq23}
\begin{array}{l}
\Phi=\bigintss_{-\frac{1}{2}}^{\frac{1}{2}} \bar{\phi}_t \ \mathrm{d} X=2 \tilde{k} \bigintss_{~G_1}^{G_2} \frac{\bar{\phi}_t}{\sqrt{F(\tilde{G})}}\  \mathrm{d} \tilde{G}.
\end{array}
\end{eqnarray}
It can be seen  that $\Phi$ depends on five quantities $\lambda_2$, $\tilde{k}, P, G_1$ and $G_2$, since the latter three can be determined from Eqs. $(\ref{Eq19})$-$(\ref{Eq21})$ for a given $\lambda_2$ for each branch (i.e., a value of $\tilde{k}$), $\Phi$ can be regarded as a function of $\lambda_2$ and $\tilde{k}$. 


\section{Bifurcation Points and Post-bifurcation Analysis}

\subsection{Bifurcation Points}
Obviously, the system Eqs. $(\ref{Eq13})$ and $(\ref{Eq16})$ always has a trivial solution. But, for such a nonlinear eigenvalue problems there are also many branches of nontrivial solutions, each of which emerges after the loading parameter $\lambda_2$ becomes larger than a critical value $\lambda_2^c$ (bifurcation point).  To determine it, we observe that near the bifurcation point both $G_1$ and $G_2$ are small. For $G_1\leq G \leq G_2$ we can approximate Eq. $(\ref{Eq19})$ and $F(\tilde{G})$ by two-term Taylor expansions. As a result, from Eqs. (\ref{Eq19})-(\ref{Eq21}) we have 
$$G_1+G_2=\frac{2 A D_1}{A_2},$$
$$2 D_1 G_1-D_2+4 (k H_0 \pi D_2)^2=0$$
and
$$\frac{2 G_1}{D_2}+(\frac{2 D_1 G_1}{D_2^2}-\frac{1}{D_2})(G_1-G_2)=0.$$
Then, by letting $G_1\rightarrow 0$, we have $G_2\rightarrow 0$, $P \rightarrow 0$ and
\begin{eqnarray}
\label{Eq24}
\begin{array}{l}
\tilde{k}^2=k^2 H_0^2=\frac{1}{ (2 \pi)^2 d_2},
\end{array}
\end{eqnarray}
where $d_2=D_2|_{P=0}$. This equation determines a critical stretch $\lambda_2^c$ for every given mode $k$.

The above simple relation provides an important insight. For a gel in a solvent, once material parameters $N\nu$ and $\chi$ (see Eq. $(\ref{Eq1})$) are specified,
there is an upper bound for the stretch  $\lambda_2$ in a homogeneous swelling. It turns out that the right-hand side of Eq. $(\ref{Eq24})$ is a decreasing function of $\lambda_2$, and as a result this upper bound for stretch $\lambda_2$ means that there is a lower bound for $k$ for a given $H_0$. Consequently, for a fixed $H_0$ some modes  will not appear naturally if their mode numbers are smaller than this lower bound. For instance, given $N\nu=10^{-3}, \chi=0.4$ and $H_0=0.01$, we have $\lambda_2 \leq 3.45$. According to Eq. (\ref{Eq24}), $k$ should be no less than $11$, which indicates that the first $10$ modes never appear. This explains why in some experiments (see \citet{HTanaka1992}) small modes were never observed.

\subsection{Bifurcation Diagrams and Post-bifurcation Analysis}
\begin{figure}
\begin{center}
\includegraphics[scale=0.8]{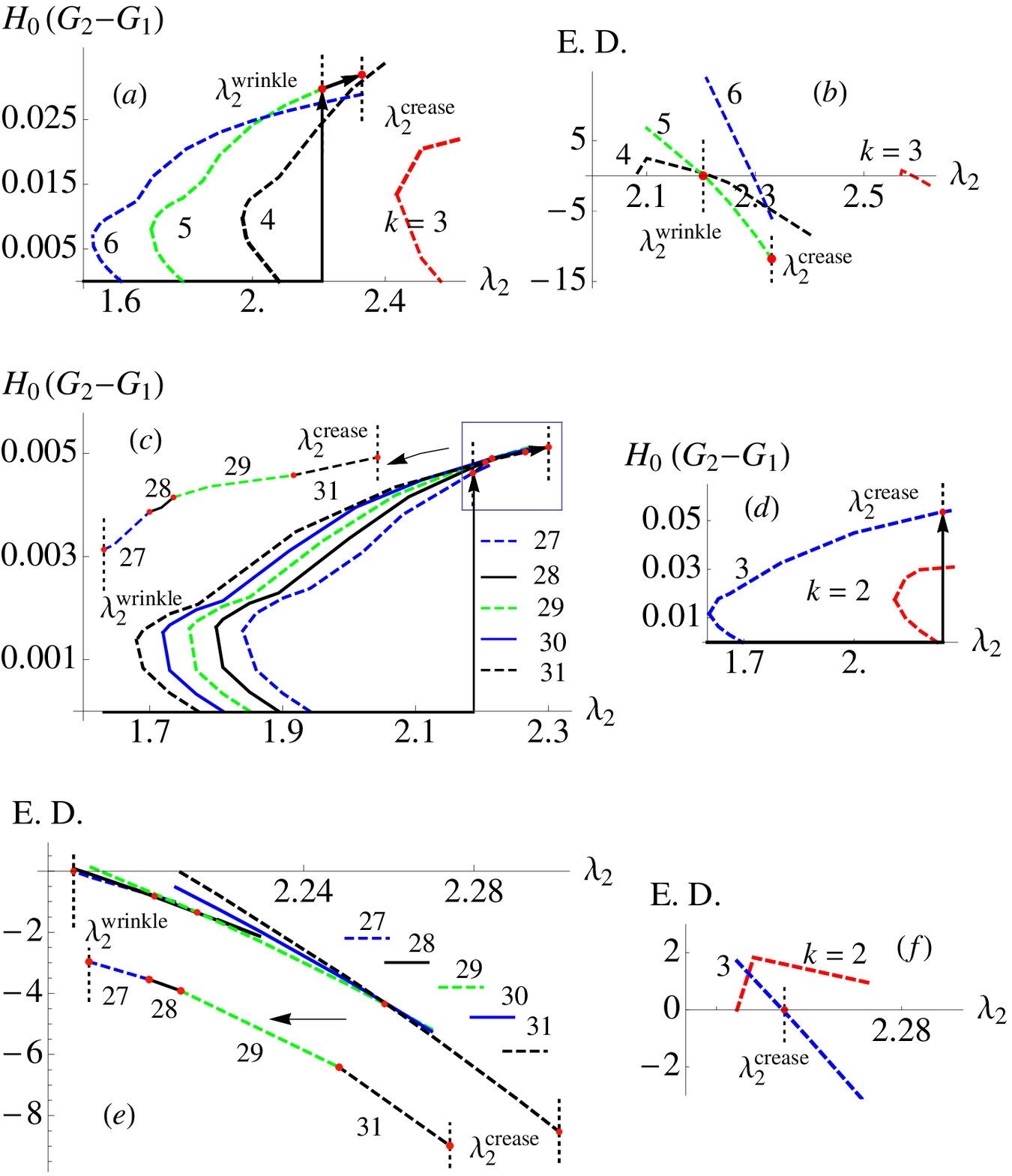}
\caption{\label{Fig_2} Bifurcation diagrams of different modes for $(a)$ $H_0=0.055$, (c) $H_0=0.0090$ and (d) $H_0=0.1$, and (b), (e) and (f) are the corresponding energy difference (E.D.) between the energy values of each bifurcated branch and the homogeneous one. For $(a)$ and $(b)$, wrinkle and crease happen at $\lambda_2^{wrinkle}=2.21$ and $\lambda_2^{crease}=2.3$ respectively both with branch $k=5$. For $(c)$ and $(e)$, wrinkle occurs at $\lambda_2^{wrinkle}=2.186$ with $k=27$, evolves and jumps to branches $k=28, 29, 31$ at $\lambda_2=2.205, 2.215, 2.265$ consecutively and finally it evolves to crease at $\lambda_2^{crease}=2.3$ with $k=31$. For (d) and (f), crease occurs directly from a flat sate at $\lambda_2^{crease}=2.3$ with $k=3$.}
\end{center}
\end{figure}

After the critical stretch for a branch is obtained, we further solve Eqs. $(\ref{Eq19})$-$(\ref{Eq21})$ for other values of $\lambda_2$ to obtain the post-bifurcation solutions. The bifurcation diagrams are shown in Figs. \ref{Fig_2}(a, c, d) for $H_0=0.055, 0.009, 0.1$ respectively. The corresponding lowest modes are respectively $k=2, 13, 2$. We can see the common characteristic shared by all bifurcation diagrams is that every branch is due to a subcritical pitchfork bifurcation followed by a saddle-node bifurcation (please note that those branches with $G_1$ as the maximum and $G_2$ as the minimum are not shown). Figs. \ref{Fig_2}$(b, e, f)$ show the energy difference (E.D.) between the energy values of the nontrivial solution and trivial one for every bifurcated branch in Figs. \ref{Fig_2}$(a,c,d)$.  We take the branch with the smallest energy value as the optimal one. It can be seen from Fig. \ref{Fig_2}(b) that for $\lambda_2<2.21$ the optimal solution is the trivial homogeneous state but after $\lambda_2>2.21$ the $k=5$ branch is the optimal one. This implies an instability taking place at $\lambda_2=2.21$, which is of a snap-through type as shown in Fig. \ref{Fig_2}$(a)$.  Of course, an instability of this nature is due to nonlinearity (if the nonlinear term $(G'(X))^2$ in Eq. $(\ref{Eq13})$ is removed, the linearized equation cannot capture the snap-through or the saddle-node bifurcation point).

Now, we examine the nature of the pattern of the optimal post-bifurcation branch, as either a wrinkle pattern or a crease pattern can occur. Here, we identify crease by using the definition in \cite{Hwa1988} and \cite{Onuki1989}: singularity arises when the slope of the surface valley goes to infinity, i.e.
$$\frac{d\tilde{y}_{surface}}{d\tilde{x}}|_{valley}=\frac{G^{\prime}(X)}{1+W'(X)}|_{valley}\rightarrow\infty,$$
where $\tilde{y}_{surface}(\tilde{x})=h_0+G(X)$ and $\tilde{x}=X+W(X)$. From Eq. $(\ref{Eq13})$, it is straightforward to prove the necessary and sufficient condition for singularity to arise is $W'(X)|_{valley}=-1$.
We denote $\lambda_2^{wrinkle}, \lambda_2^{crease}$ as the stretches where wrinkle or crease just occur. Upon using $W'(X)=-1|_{valley}$ and $G'(X)|_{valley}=0$ in Eq. (\ref{Eq13}), we obtain
\begin{equation}
\label{Eq25}
0=f_4(P)G_1 - \lambda_2(E_2+f_2(P)(1+f_6(P)).
\end{equation}
This equation together with Eqs. $(\ref{Eq19})$-$(\ref{Eq21})$  can be used to find $\lambda_2^{crease}$ for each branch and the corresponding values of $G_1, G_2, P$ at the crease point.

\begin{figure}
\begin{center}
\includegraphics[scale=0.65]{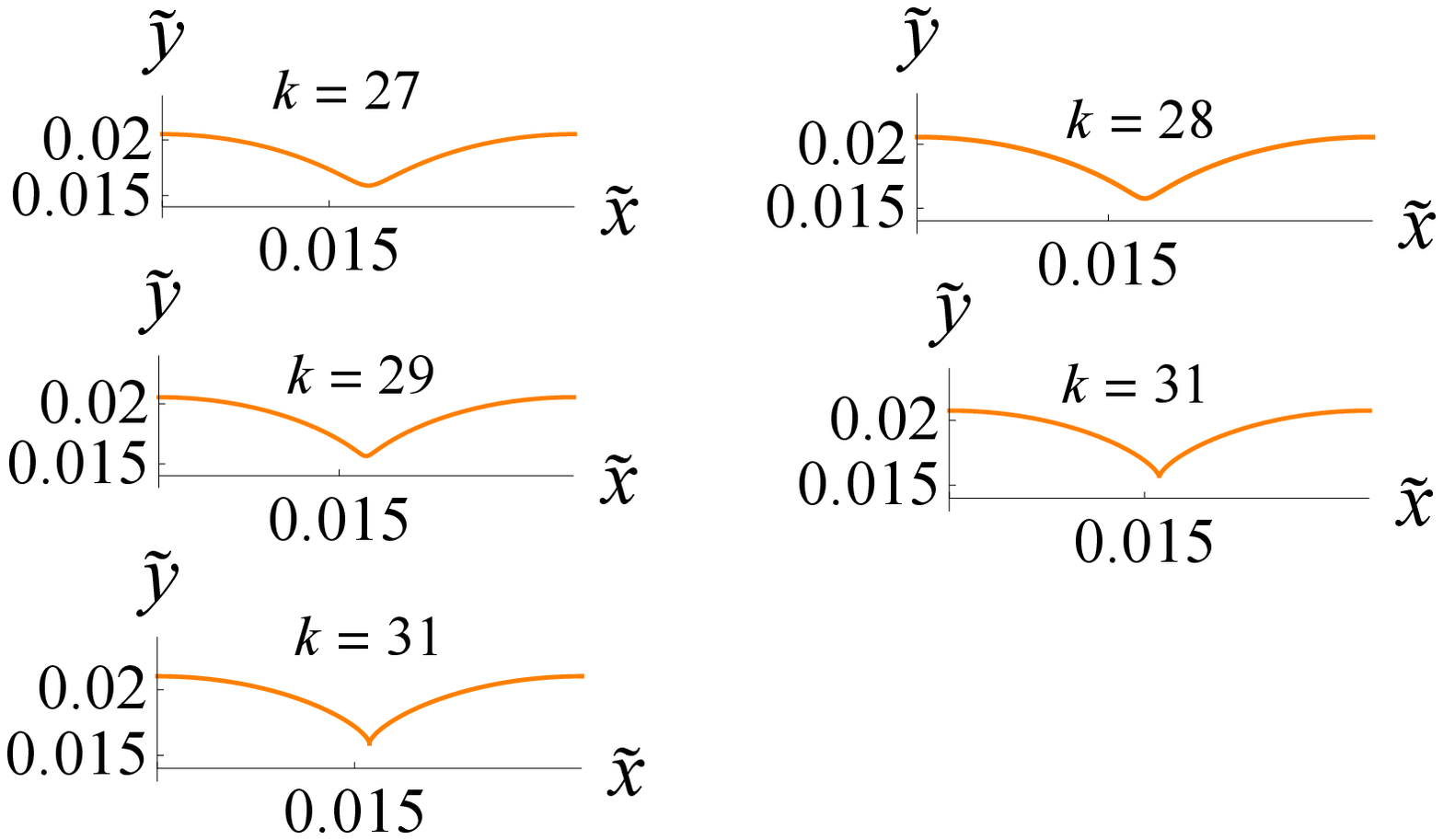}
\caption{\label{Fig_3} Surface profile evolution (in one period) from wrinkle to crease through modes jump for $H_0=0.009$. $\lambda_2$ values: $(a)$ $2.186$, (b) $2.205$, (c) $2.215$, (d) $2.265$, (e) $2.3$.}
\end{center}
\end{figure}

It is found that for $H_0=0.055$ (Fig. \ref{Fig_2}(b)), $\lambda_2^{wrinkle}=2.21$ and the optimal $k=5$ branch gives a wrinkle pattern. As $\lambda_2$ further increases, $W'(X)|_{valley}$ moves towards $-1$, and for this branch we find $\lambda_2^{crease}=2.3$, at which a crease pattern occurs. This is a pathway from wrinkle to crease (called set-back crease in \citet{Diab2013}).  Fig.\ref{Fig_2}(e) ($H_0=0.009$) shows another pathway to crease through modes jumps (which was observed experimentally by \cite{HTanaka1992} but was never studied theoretically or even numerically). From Fig. \ref{Fig_2}(e), we can see that the gel layer is flat until $\lambda_2$ reaches 2.186 ($\lambda_2^{wrinkle}$) where the optimal branch with $k=27$ emerges by a snap-through (see Fig. \ref{Fig_2}(c)). Then, as $\lambda_2$ further increases the optimal branch switches consecutively to $k=28, 29, 31$ branches. For $k=27, 28, 29, 31$ and $\lambda_2<2.3$, those are wrinkle patterns. It is at  $k=31$ branch and when $\lambda_2^{crease}=2.3$ is reached that a crease pattern occurs. A third pathway to crease is shown in Figs. \ref{Fig_2}(d, f) ($H_0=0.1$).  At $\lambda_2=2.3$ ($\lambda_2^{crease}$) the gel layer jumps from a flat state directly to a crease state with $k=3$. In Fig. \ref{Fig_3}, we show the evolution of the surface profile in one period for $H_0=0.009$, for which the pathway to crease is through mode jumps.

In summary, depending on the thickness of the layer, there are three pathways to crease. In the first two pathways wrinkle appears first after a snap-through and then crease occurs in the same branch or another branch through mode jumps. We do not find definite thickness intervals to distinguish these two cases. But, in general for a very thin layer (say, $H_0\leq 0.036$) mode jumps happen before crease occurs. The third pathway from a homogeneous state directly to crease happens for $H_0\geq 0.096$. We also mention that the bifurcations (leading to wrinkle or crease) occur for $\lambda_2$ in a small interval $[2.15, 2.3]$, which may explain the discrepancy between the theory here and some experiments where wrinkle was not observed before crease.

\subsection{Interpretations of some Experimental Results}
Several experiments on gel layers swellings have been conducted \citep{HTanaka1992,HTanaka1994,Trujillo2008}. The gels used in those works may not be modeled by the free energy density Eq. $(\ref{Eq1})$. Nevertheless, qualitatively several features observed in those experiments are well captured by our analytical results and can be interpreted nicely.  

\citet{HTanaka1992} and \citet{Trujillo2008} observed that the wavelength in a crease pattern is proportional to the initial layer thickness ($H_0$). And, in \cite{Trujillo2008} a critical strain at the moment of crease formation was introduced, which was defined as $\epsilon^{crease}=1-L/L^f$ where $L^f$ is the layer length if the gel is allowed to swell freely under the same chemical condition. It was found that $\epsilon^{crease}$ is independent of $H_0$. Now, we give the explanations of those two results. For each bifurcated branch, the total energy $\Phi$ is a function of $\tilde{k}$ and $\lambda_2$, see Eqs. (\ref{Eq23}) (which was obtained upon using the derived scaling laws). For the optimal branch, the selection of $\tilde{k}$ is made according to 
\begin{equation}
\label{Eq26}
\frac{\partial \Phi}{\partial \tilde{k}}= 0.
\end{equation}
This equation together with  Eqs. (\ref{Eq19})-(\ref{Eq21}) and (\ref{Eq25}) determine the five quantities $\lambda_2^{crease}$, $\tilde{k}$, $P$, $G_1$ and $G_2$ completely, implying that all of them are independent of $H_0$. Therefore, $\tilde{k}=constant$ implies that wavelength $\frac{1}{k}=constant*H_0$, which explains the result by \cite{HTanaka1992}.  For a value of $\lambda_2^{crease}$, one can find the corresponding value of $\mu$, which, in turn, can be used to find $L_f$, the length of the layer in a free swelling. Then, $\epsilon^{crease}$ is also determined. Since $\lambda_2^{crease}$ is invariant with respect to $H_0$, so is $\epsilon^{crease}$, and this explains of  the result found in \cite{Trujillo2008}.

\begin{figure}
\begin{center}
\includegraphics[scale=0.7]{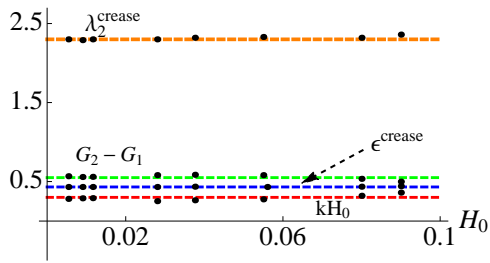}
\caption{\label{Fig_4} Curves of four physical quantities versus the initial thickness where $\lambda_2^{crease}=2.3, G_2-G_1=0.55, \epsilon^{crease}=0.43$ and $k H_0=0.3.$}
\end{center}
\end{figure}

At the moment of crease formation, $G_2-G_1$ is also constant. Then, the scaling above Eq. (\ref{Eq17}) implies that the amplitude ($=H_0(G_2-G_1)$) of the crease profile is proportional to $H_0$. And, further $k H_0(G_2-G_1)$, which is the ratio of amplitude-to-wavelength (a representation of the crease shape), is constant. In this sense, the crease shape is independent of the thickness. In Fig. \ref{Fig_4}, the quantities $kH_0$, $\epsilon^{crease}$, $\lambda_2^{crease}$ and $G_2-G_1$ are also plotted, whose independence on $H_0$ can be clearly seen. We note that those constant values in Fig. \ref{Fig_4} depend on the chosen values of $N\nu$ and $\chi$. Nevertheless, for the chosen $N\nu$ and $\chi$, $\lambda_2^{crease}=2.3$ is close to the value $2.4$ found in \cite{Hong2009} experimentally and numerically for creases in elastomers. In the experiment by \cite{Guvendiren2010}, it was found that the increase of the density of the polymer cross-link ($N\nu$) reduced the critical stretch for crease, and this effect is also captured by our analytical results. We find that when $N\nu$ increases from $10^{-3}$ to $10^{-2}$, $\lambda_2^{wrinkle}$ decreases from $2.2$ to $1.62$ while $\lambda^{crease}$  decreases  from $2.3$ to $1.63$. It should be noted that the derived reduced system is valid up to the onset of crease formation, and the crease evolution afterwards cannot be addressed by the present theory.

\section{Conclusion}
Crease formation is a widely-spread phenomenon. Despite of many experimental and numerical studies, our understanding on this phenomenon is still limited. With an aim of offering a more comprehensive picture leading to crease formation, here we carry out a theoretical investigation on instabilities in a swelling gel layer based on analytical methods. The success relies on using a method of coupled series-asymptotic expansions developed by one of the authors earlier to reduce the original eignevalue problem of nonlinear PDEs to that of nonlinear ODEs. Closed-form solutions are constructed for the latter problem, and thus all post-bifurcation modes are captured, which is one of the unique features of the present analytical study. As a consequence, deep insights are provided, including (a)determination of all bifurcation points and the bifurcation type; (b)establishing a lower bound for mode numbers; (c)identifying the optimal bifurcated branch; (d)unveiling three pathways to crease formations; (e)deriving two scaling laws.  A number of interesting experimental features are captured qualitatively and interpreted. Based on the derived system, a set of five algebraic equations for determining physical quantities at the crease point is also provided, which reveals that some basic quantities are invariant with respect to the layer thickness. In our view, for snap-through instability phenomena governed by nonlinear eigenvalue problems, analytical results may be indispensable for a complete description, as demonstrated here for crease formations in a gel layer.

\section*{References}


\end{document}